\documentclass{article}
\usepackage{hiph-art}
\usepackage{graphicx}
\volnumber{19} \issuenumber{1} \edyear{2004}                             
\frompage{000} \topage{000}                                              
\recrevdate{15 April 2004}                                               
\newcommand{\Aref}[1]{$^{#1}$}
\newcommand{\etal}{{\it et~al.}} 
\newcommand{\mT}{m_{\mathrm T}}
\newcommand{\pT}{p_{\mathrm T}}

\newcommand{\dder}{{\mathrm{d}}}

\def\rightmark{Results on hyperon production from the NA57 experiment} 
\def\leftmark{Ladislav~\v{S}\'{a}ndor for the NA57 collaboration}

\title{Results on hyperon production from the NA57 experiment} 
\authors{
  Ladislav~\v{S}\'{a}ndor for the NA57 collaboration: \\[2ex]
  {\footnotesize
  F.~Antinori\Aref{k},
  P.A.~Bacon\Aref{e},
  A.~Badal\`{a}\Aref{f},
  R.~Barbera\Aref{f},
  A.~Belogianni\Aref{a},
  I.J.~Bloodworth\Aref{e},
  M.~Bombara\Aref{h},
  G.E.~Bruno\Aref{b},
  S.A.~Bull\Aref{e},
  R.~Caliandro\Aref{b},
  M.~Campbell\Aref{g},
  W.~Carena\Aref{g},
  N.~Carrer\Aref{g},
  R.F.~Clarke\Aref{e},
  A.~Dainese\Aref{k},
  D.~Di~Bari\Aref{b},
  S.~Di~Liberto\Aref{n},
  R.~Divia\Aref{g},
  D.~Elia\Aref{b},
  D.~Evans\Aref{e},
  G.A.~Feofilov\Aref{p},
  R.A.~Fini\Aref{b},
  P.~Ganoti\Aref{a},
  B.~Ghidini\Aref{b},
  G.~Grella\Aref{o},
  H.~Helstrup\Aref{d},
  K.F.~Hetland\Aref{d},
  A.K.~Holme\Aref{j},
  A.~Jacholkowski\Aref{f},
  G.T.~Jones\Aref{e},
  P.~Jovanovic\Aref{e},
  A.~Jusko\Aref{e},
  R.~Kamer\-mans\Aref{r},
  J.B.~Kinson\Aref{e},
  K.~Knudson\Aref{g},
  V.~Kondratiev\Aref{p},
  I.~Kr\'alik\Aref{h},
  A.~Krav\v{c}\'{a}kov\'{a}\Aref{i},
  P.~Kuijer\Aref{r},
  V.~Lenti\Aref{b},
  R.~Lietava\Aref{e},
  G.~L\o vh\o iden\Aref{j},
  V.~Manzari\Aref{b},
  M.A.~Mazzoni\Aref{n},
  F.~Meddi\Aref{n},
  A.~Michalon\Aref{q},
  M.~Moran\-do\Aref{k},
  P.I.~Norman\Aref{e},
  A.~Palmeri\Aref{f},
  G.S.~Pappalardo\Aref{f},
  B.~Pastir\v{c}\'ak\Aref{h},
  R.J.~Platt\Aref{e},
  E.~Quercigh\Aref{k},
  F.~Riggi\Aref{f},
  D.~R\"{o}hrich\Aref{c},
  G.~Ro\-mano\Aref{o},
  K.~\v{S}afa\v{r}\'{\i}k\Aref{g},
  L.~\v{S}\'andor\Aref{h},
  E.~Schillings\Aref{r},
  G.~Segato\Aref{k},
  M.~Sen\'{e}\Aref{l},
  R.~Sen\'{e}\Aref{l},
  W.~Snoeys\Aref{g},
  F.~Soramel\Aref{k}\Aref{*},
  M.~Spyropoulou-Stassinaki\Aref{a},
  P.~Staroba\Aref{m},
  R.~Turrisi\Aref{k},
  T.S.~Tveter\Aref{j},
  J.~Urb\'{a}n\Aref{i},
  P.~van~de~Ven\Aref{r},
  P.~Vande~Vyvre\Aref{g},
  A.~Vascotto\Aref{g},
  T.~Vik\Aref{j},
  O.~Villalobos Baillie\Aref{e},
  L.~Vinogradov\Aref{p},
  T.~Virgili\Aref{o},
  M.F.~Vot\-ruba\Aref{e},
  J.~Vrl\'{a}kov\'{a}\Aref{i} and
  P.~Z\'{a}vada\Aref{m}
}
\scriptsize\begin{flushleft}
  $^{a}$  Physics Department, University of Athens, Athens, Greece \\
  $^{b}$  Dipartimento I.A. di Fisica dell'Universit\`{a} e del
    Politecnico di Bari and INFN, Bari, Italy \\
  $^{c}$  Fysisk institutt, Universitetet i Bergen, Bergen, Norway \\
  $^{d}$  H\o gskolen i Bergen, Bergen, Norway \\
  $^{e}$  School of Physics and Astronomy, University of Birmingham,
    Birmingham, UK \\
  $^{f}$  Dipartimento di Fisica dell'Universit\`{a} and Sezione INFN,
    Catania, Italy \\
  $^{g}$  CERN, European Laboratory for Particle Physics, Geneva,
    Switzerland \\
  $^{h}$  Institute of Experimental Physics, Slovak Academy of Science,
    Ko\v{s}ice, Slovakia \\
  $^{i}$  P.J. \v{S}af\'{a}rik University, Ko\v{s}ice, Slovakia \\
  $^{j}$  Fysisk institutt, Universitetet i Oslo, Oslo, Norway \\
  $^{k}$  Dipartimento di Fisica dell'Universit\`{a} and Sezione
    INFN, Padua, Italy \\
  $^{l}$  Coll\`{e}ge de France and IN2P3, Paris, France \\
  $^{m}$  Institute of Physics, Academy of Science of the Czech Republic,
    Prague, Czech Republic \\
  $^{n}$  Dipartimento di Fisica dell'Universit\`{a} ``La Sapienza''
    and Sezione INFN, Rome,Italy \\
  $^{o}$  Dipartimento di Scienze Fisiche ``E.R. Caianiello''
    dell'Universit\`{a} and INFN, Salerno, Italy \\
  $^{p}$  State University of St. Petersburg, St. Petersburg, Russia \\
  $^{q}$  Institut de Recherches Subatomiques, IN2P3/ULP, Strasbourg,
    France \\
  $^{r}$  Utrecht University and NIKHEF, Utrecht, The Netherlands\\
  $^{*}$  Permanent address: University of Udine, Italy
\end{flushleft}}
 
\abstract{Recent results on hyperon production in 
Pb--Pb collisions from the NA57 experiment are reported.
Strangeness enhancements and the transverse mass 
spectra properties at 158 GeV per nucleon are described.}
 
\begin{document}
 
\maketitle

\section{Introduction}
The extensive study of ultrarelativistic heavy-ion collisions in the past two decades
was motivated mainly by the QCD prediction that at sufficiently 
high energy density excited matter undergoes a phase transition to a system of 
deconfined quarks and gluons (quark-gluon plasma, QGP). Strange particle production 
has proven to be a powerful tool for the investigation of the dynamics 
of heavy-ion collisions at high energies (see, e.g. \cite{sqm03}).

In particular, the WA97 experiment at the CERN SPS has measured an enhanced 
production of strange and multi-strange hyperons in central Pb--Pb collisions at 158
$A$ GeV/$c$ with respect to proton induced reactions \cite{wa97}. The observed pattern
of the
enhancements, increasing with the strangeness content of the particle, was predicted by
Rafelski and M\"{u}ller \cite{RaMu} more than 20 years ago as a consequence of a QGP 
formation.

NA57, a second generation SPS heavy-ion experiment, continues and extends the
study of the production of
strange and multi-strange particles in Pb--Pb collisions initiated by the 
WA97 experiment, collecting data in wider centrality range and at two beam momenta - 40
and 158 $A$ GeV/$c$. In this paper, based on a brief review of recent NA57 results
given at the 3rd Budapest winter school on heavy-ion physics in December 2003, we
concentrate on strangeness enhancement measurements and on the analysis of the 
transverse mass
spectra of hyperons from Pb--Pb collisions at 158 $A$ GeV/$c$.     
   
\section{The NA57 experiment}
The NA57 set-up and experimental procedure are described elsewhere 
\cite{setup_1,setup_2,ex_proc}. 
A telescope of compactly packed high granularity silicon pixel 
detectors, located in a \mbox{1.4~T} magnetic field, is used as 
main tracking device. The configuration and geometry of telescope enable
to measure particle yields and spectra over half a unit of rapidity
around the  mid-rapidity. The centrality trigger, based on an array of scintillator 
counters, selects the 60\% most central fraction of the Pb--Pb inelastic cross 
section. Two stations of silicon microstrip detectors provide data for the measurement
of the charged particle multiplicity used for the
offline collision centrality determination. As a measure of centrality we use 
the number of wounded nucleons computed from the measured trigger cross sections
via the Glauber model \cite{nicola}.

NA57 has collected Pb--Pb data samples at both 158 and 40 $A$ GeV/$c$ beam momentum.
In order to study the strangeness enhancements at
both energies the knowledge of hyperon production in elementary hadron processes
is necessary. For this purpose NA57 has also collected a reference data sample of 
\mbox{p--Be} collisions at 40 $A$ GeV/$c$. At 158 $A$ GeV/$c$ 
the p--Be reference data available from the WA97 experiment were utilized.

The $\mathrm{K}^0_S$ mesons, the $\Lambda,~\Xi^-$ and $\Omega^-$ hyperons and their 
antiparticles
are identified by reconstructing their weak decays into final states containing 
charged particles only. The selection procedure allows us to extract strange particle
signals with \mbox{a~negligible} level of background. Data are corrected for 
geometrical acceptance and for detector and reconstruction inefficiencies.
We have used all the reconstructed multi-strange particles, while for the 
much more abundant $\mathrm{K}^0_S,~\Lambda$ and $\overline\Lambda$ particles we 
only corrected a fraction
of the total data samples in order to reach a statistical accuracy better than 
the limits imposed by the systematics. The systematic errors for both particle
yields and spectra are estimated to be about $10\%$ for $\Lambda,~\Xi$ and 
$\simeq 15\%$ for $\Omega$ hyperons.

The double differential cross sections for each particle under study were fitted 
using the expression
  \begin{equation}
   \frac{\dder^2 N}{\dder\mT\,\dder y} 
    = f(y)\,\mT\,\exp\left(-\frac{\mT}{T_{app}}\right)~,
  \label{ymtdist}
  \vspace{0.15cm}
  \end{equation}
where $\mT=\sqrt{m^2+p_{\mathrm{T}}^2}$  is the transverse mass and $T_{app}$ is
the inverse slope para\-meter, 
assuming the rapidity 
distribution to be flat in our limited acceptance region ($f(y)$ = const.).
  
\begin{figure}[b]
\vspace{0.15cm}
\centering
 \resizebox{0.475\textwidth}{!}{%
 \includegraphics{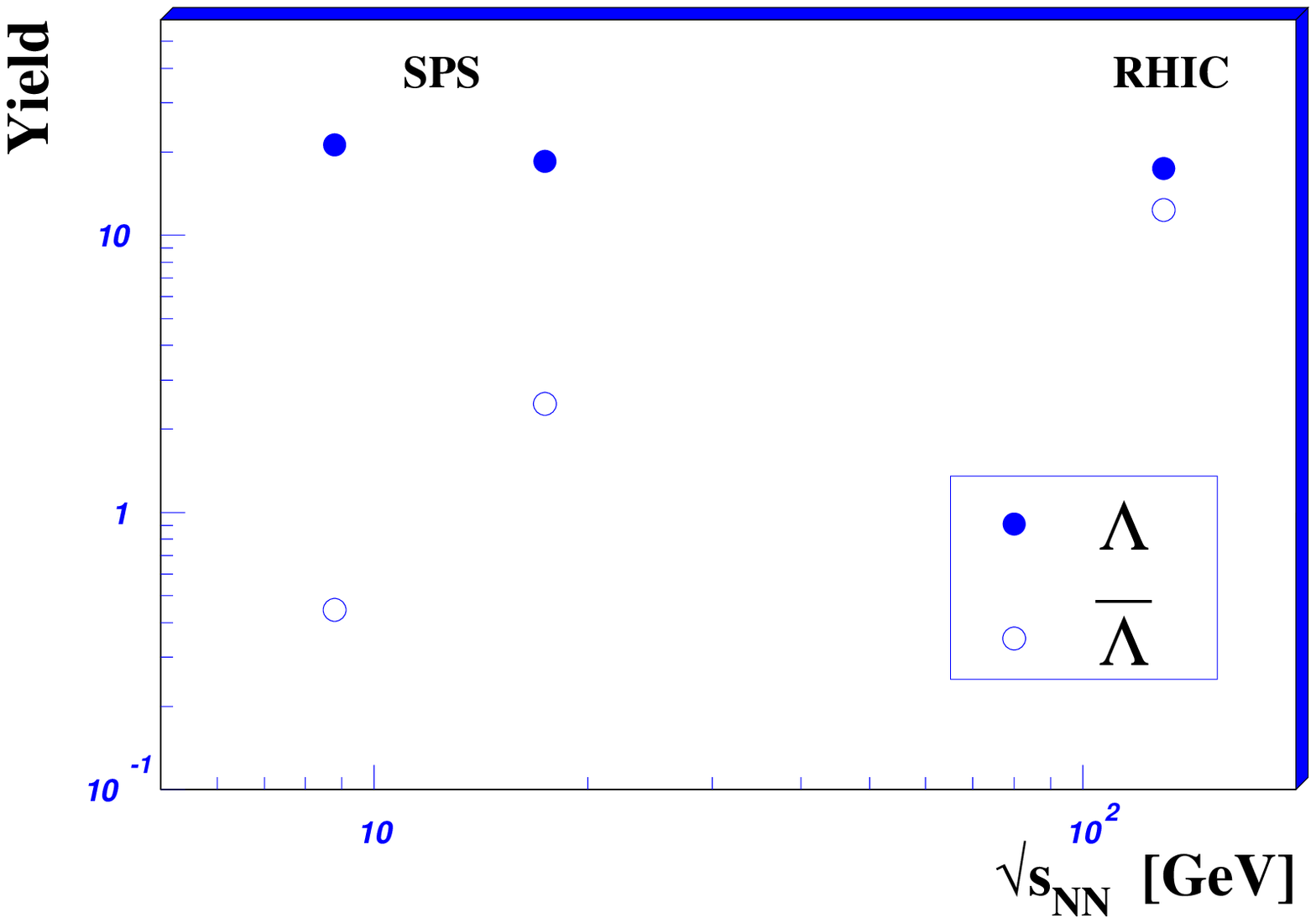}}
 \hspace{-0.49cm}
 \resizebox{0.475\textwidth}{!}{%
 \includegraphics{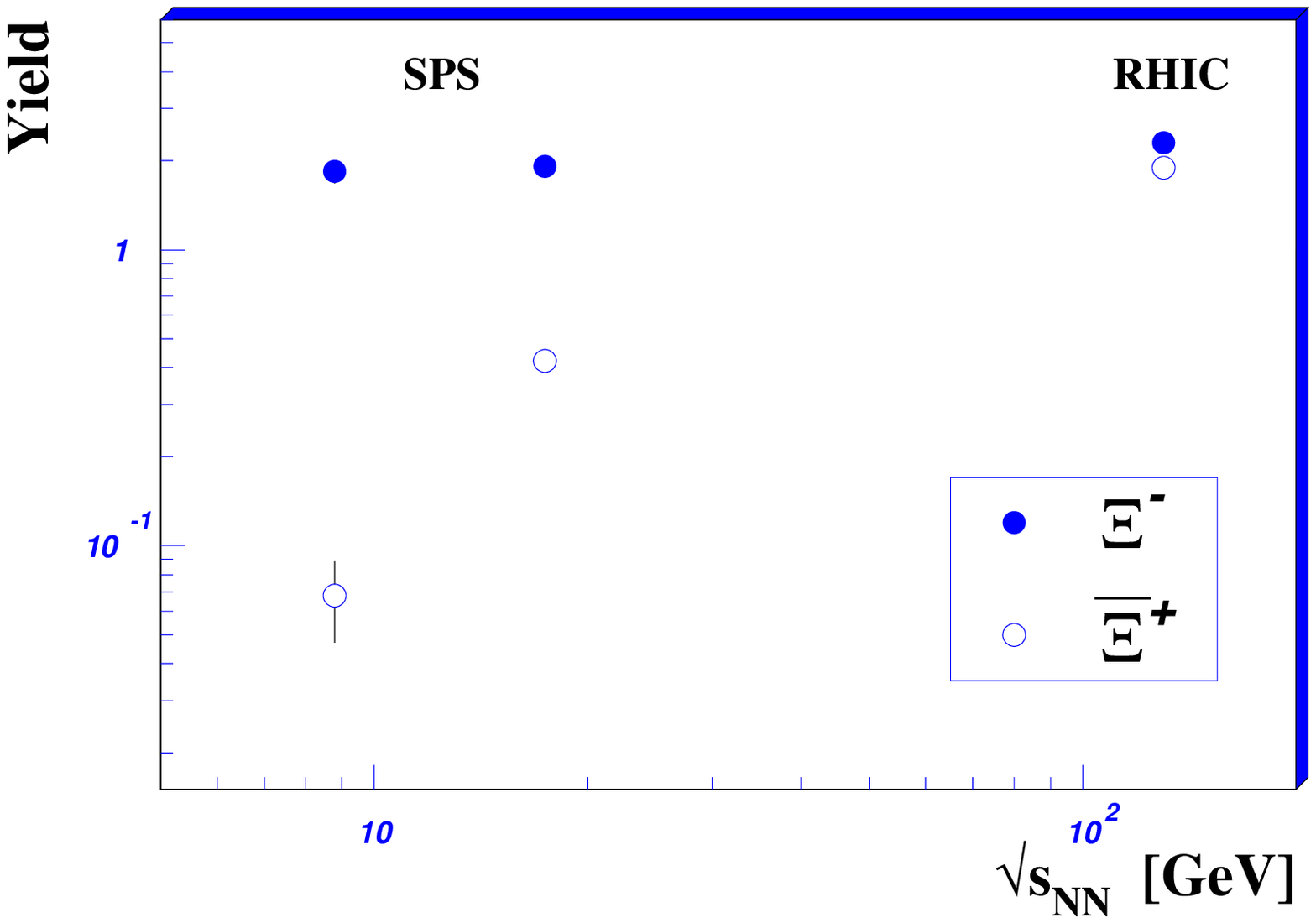}}
 \resizebox{0.475\textwidth}{!}{%
 \includegraphics{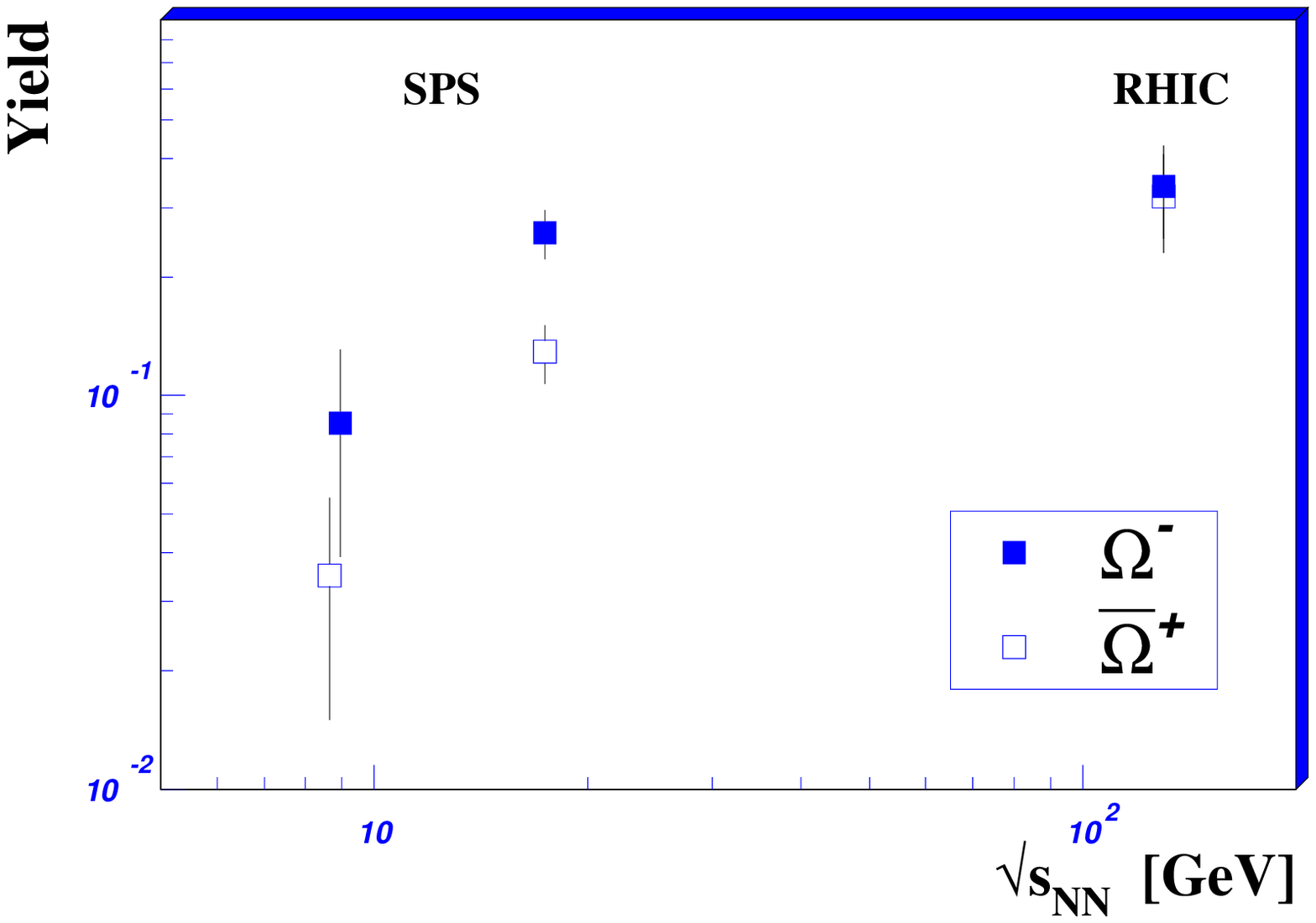}}
\vspace{-0.25cm}
\caption{\label{yields_na57_rhic} \small Energy dependence of hyperon yields $Y$ in 
         central heavy-ion collisions.}
\vspace{0.15cm}
\end{figure}
 
\section{Particle yields and strangeness enhancements}

Using the parametrization given by equation (\ref{ymtdist}), with the
$T_{app}$ value extracted from from the maximum likelihood fit to the data, one 
can determine 
the yield (number of particles per rapidity unit centred at mid-rapidity $y_0$) 
of each particle under study:
\vspace{0.20cm}
\begin{equation}
Y = \int^{\infty}_m \!\!\dder\mT 
\int^{y_{0}+0.5}_{y_{0}-0.5} \! 
\frac{\dder^2 N}{\dder\mT\,\dder y} ~\dder y~.
\label{ex_yield}  
\vspace{0.20cm}
\end{equation}
The data have been divided into five centrality classes
(0--4) \footnote{The centrality
classes (1--4) correspond to the four classes used in the WA97 analysis,
while the most peripheral \mbox{class 0} (with
$\langle N_{\mathrm{wound}}\rangle = 58 \pm 4$) is accessible to NA57 only. The class 
4 corresponds to the 5\% and classes 3+4 to the 11\% most central fractions of
events respectively.}
and particle yields $Y$ were calculated for each centrality class.


In figure \ref{yields_na57_rhic} the NA57 yields of hyperons from central
Pb--Pb interactions at 40 and 158 $A$ GeV/$c$ are presented and compared
with those obtained by the STAR collaboration in 
$\sqrt{s_{\mbox{\tiny\it NN}}} = 130~\mathrm{GeV}$
Au--Au collisions at RHIC \cite{lam_130,cast_02,suire_02}.
In order to compare data close in centrality with those of STAR (5\%, 10\% and 11\%
most central collisions for $\Lambda,~\Xi$ and $\Omega$) we have selected NA57 data
for 5\% most central events for $\Lambda$ and 11\% most central events for $\Xi$ 
and $\Omega$. The $\Lambda$ and $\Xi^-$ yields are almost constant
in the full considered energy range, the $\Omega^-$ yield increases
with the energy. All anti-hyperon yields show a steep energy dependence.
\vspace{0.20cm}
\begin{figure}[h]
\centering
  \resizebox{0.96\textwidth}{!}{%
 \includegraphics{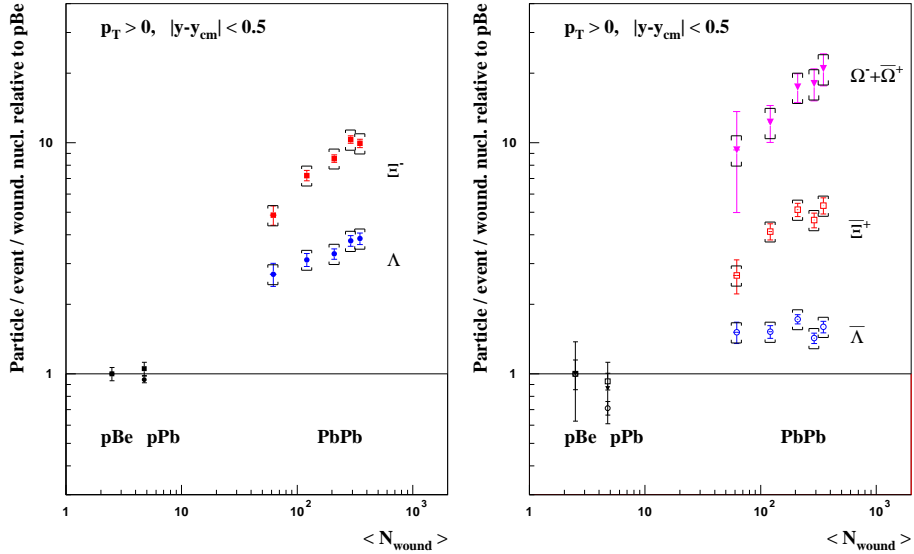}}
 \vspace{-0.2cm}
 \caption {\label{enh_158} \small Hyperon enhancements {\it E} at 158 $A$ GeV/$c$
           as a function of the number
           of wounded nucleons. The symbol $^\sqcap_\sqcup$ indicates the systematic
           error.}
\end{figure}
\vspace{-0.3cm}

Let us turn now to strangeness enhancements. Having measured the mid\-rapidity 
particle yields $Y$
for both Pb--Pb and p--Be collisions, one can determine \mbox{the~strangeness} 
enhancement
\vspace{0.20cm}
\begin{equation}
 E = \left(\frac{Y}{\langle {N_{\mathrm{wound}}} \rangle } \right)_{\mathrm{Pb-Pb}}/
     \left(\frac{Y}{\langle {N_{\mathrm{wound}}} \rangle } \right)_{\mathrm{p-Be}}~.
\label{enhance}  
\end{equation}
The NA57 results on strangeness enhancements at 158 $A$ GeV/$c$ are shown in 
\mbox{fig. \ref{enh_158}}.

The enhancements are shown separately for particles containing at least one valence
quark in common with the nucleon (left) and for those with no valence quark in common 
with the nucleon (right).
Our results confirm the pattern of strangeness enhancements observed by the WA97 
experiment \cite{wa97}. 
The enhancement increases with the strangeness content of the hyperon. 
The Pb--Pb data exhibit a significant centrality dependence of the yields per wounded
nucleon for all hyperons except for $\overline\Lambda$. However, for the two most 
central classes 3 and 4 ($\simeq 10\%$ most central fraction of collisions) 
a saturation 
of the enhancements is not excluded \footnote{At the time of this manuscript's 
preparation the yields of $\Lambda, \overline\Lambda$ and $\Xi^-$ hyperons from p--Be 
at 40 $A$ GeV/$c$ have been obtained and strangeness enhancements
in Pb--Pb data presented for the first time \cite{qm_04}. The strangeness enhancement 
pattern at 40 $A$ GeV/$c$ exhibits similar properties and follows the same hierarchy 
as observed at 158 $A$ GeV/$c$.}.
 
\section{Transverse mass spectra}

\begin{figure}[h!]
\vspace{-0.20cm}
\centering
  \resizebox{0.33\textwidth}{!}{%
 \includegraphics{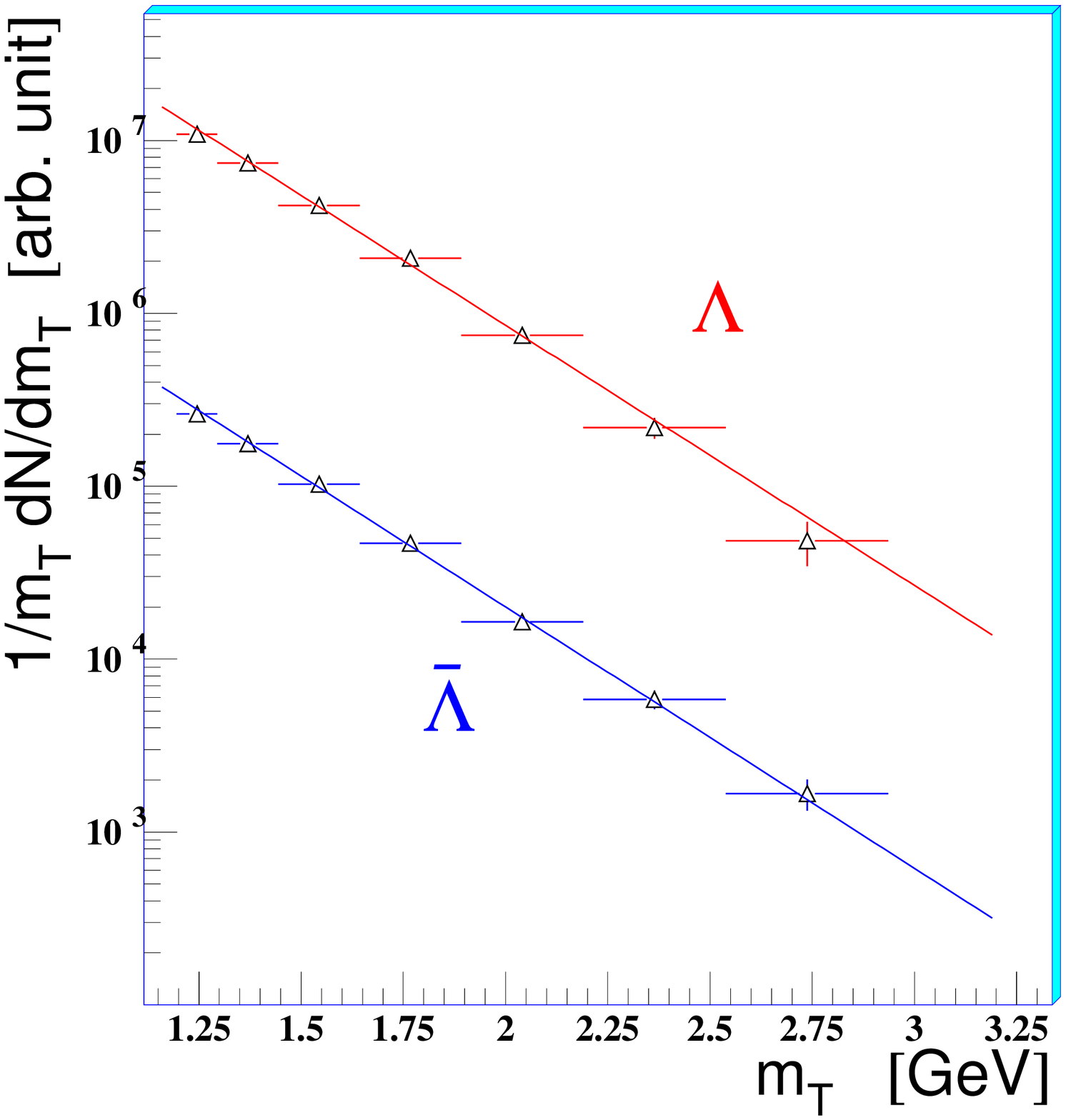}}
  \resizebox{0.305\textwidth}{!}{%
 \includegraphics{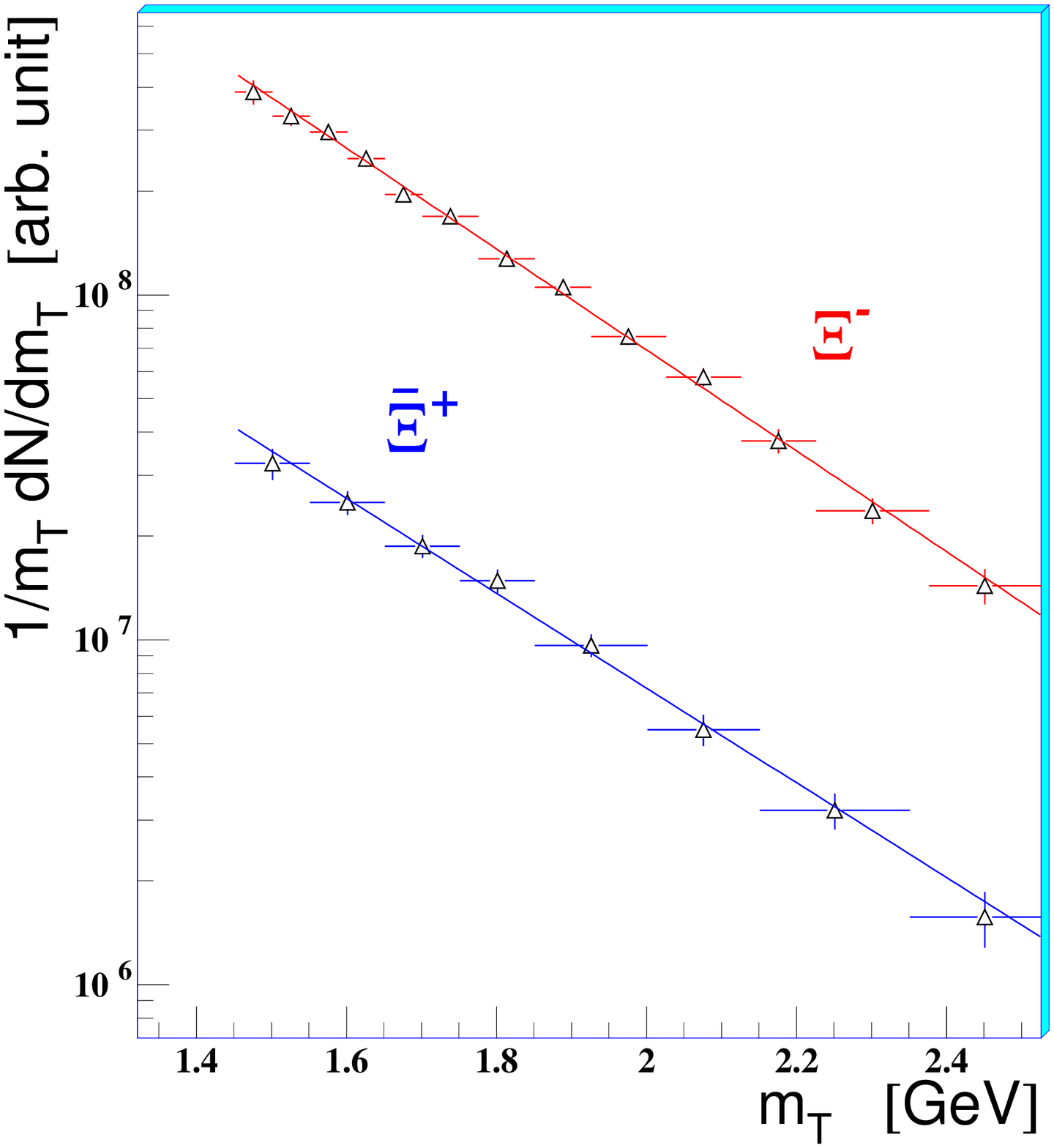}}
  \resizebox{0.33\textwidth}{!}{%
 \includegraphics{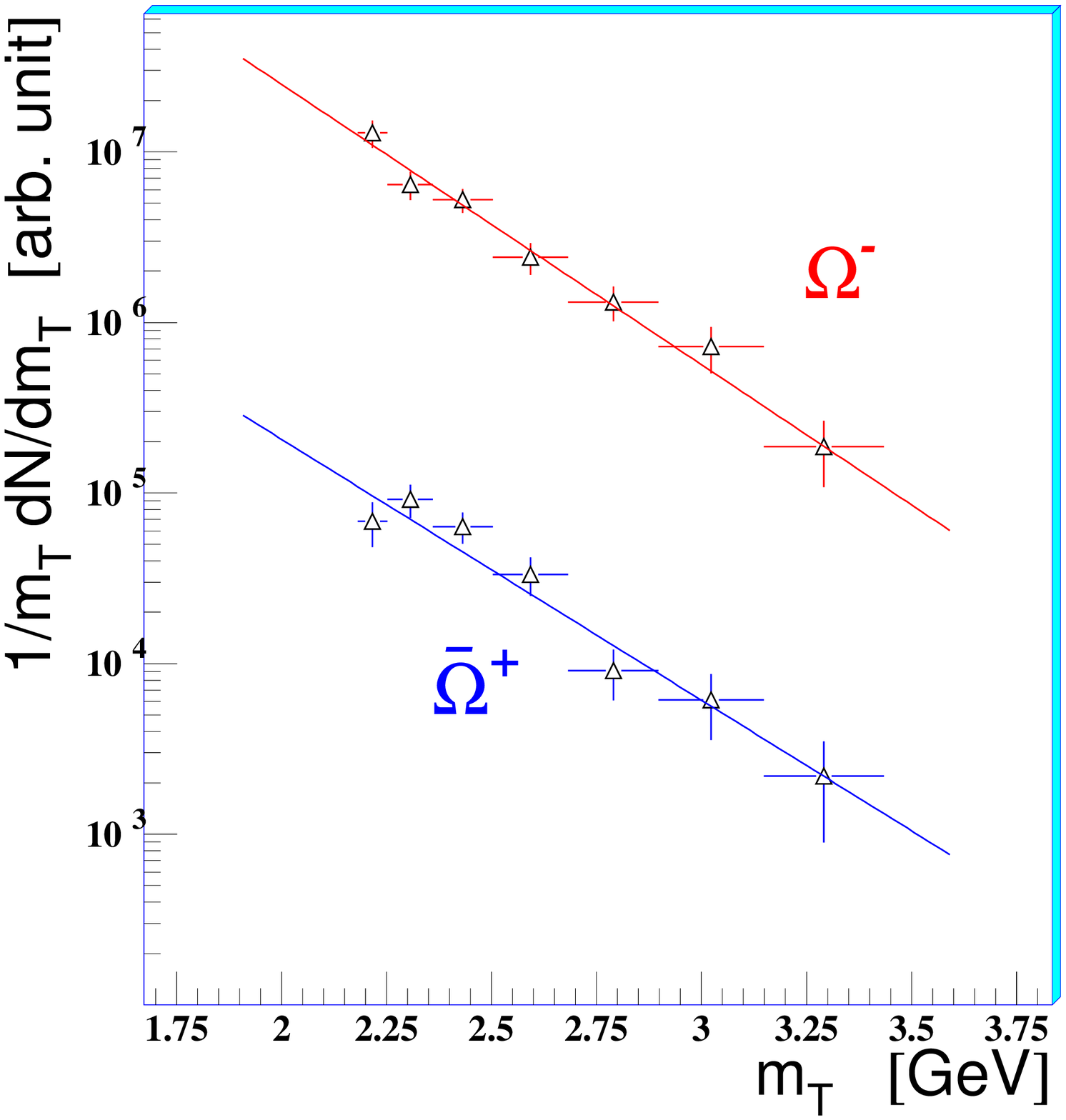}}
\vspace{-0.20cm}
\caption {\label{mt_sp158} \small Transverse mass spectra of hyperons from 158 
          $A$ GeV/$c$ Pb--Pb collisions.}
\vspace{-0.30cm}
\end{figure}

The transverse mass distributions ~$1/\mT ~\dder N/\dder \mT$~ for hyperons and 
antihyperons from Pb--Pb collisions at 158 $A$ GeV/$c$ measured in the whole
centrality range accessible to the experiment are shown in figure \ref{mt_sp158}.

The inverse slopes of the superimposed exponential functions correspond to the 
$T_{app}$ (`apparent temperature')
values extracted from the maximum likelihood fit of equation (\ref{ymtdist}) to 
the data. The shapes of all spectra are close to exponential fun\-ctions with similar 
$T_{app}$ values ($\sim 300$ MeV) for all hyperons and antihyperons \footnote{
The inverse slope values $T_{app}$ are in agreement within the errors with those 
measured
over a narrower centrality range by the WA97 experiment \cite{wa97_mt}.}. The 
observed systematics of inverse slopes $T_{app}$ is usually interpreted as 
due to the combined contributions 
of thermal motion in the fireball and collective transverse flow.

For a more complex analysis aiming at disentangling the radial flow velocity 
$\beta_\perp(r)$
and the thermal freezout temperature $T$
we have utilized a model based on thermalization and a hydrodynamical description
of transverse flow \cite{uli_93}. In this approach (the blast-wave model) the 
$\mT$ distribution of hadron $i$ can be approximated as follows:
\begin{equation}
\frac{\dder N_i}{\mT \,\dder \mT} \propto \mT \int^R_0 r\,\dder r 
K_1\left(\frac{\mT \cosh\rho}{T}\right)
I_0\left(\frac{\pT \sinh\rho}{T}\right),
\label{blast}
\end{equation}
where $R$ is the transverse system size, $K_1$ and $I_0$ are modified Bessel 
functions and $\rho={\rm atanh}\;\beta_\perp(r)$ is a transverse boost. The model
assumes kinetic freeze-out of matter at constant temperature $T$ and a cylindrically
symmetric and longitudinally boost invariant expansion. The transverse velocity
field is parametrised according to a power law
\begin{equation}
\beta_\perp(r) = \beta_s \left( \frac{r}{R} \right)^n.
\label{power}
\end{equation}
Assuming a uniform particle density the average transverse velocity is related to 
the surface velocity $\beta_s$ by the formula
$\langle \beta_\perp \rangle = 2/(2+n) \times \beta_s$.

\begin{figure}[b!]
\vspace*{-.15cm}
\centering
  \resizebox{0.45\textwidth}{!}{%
 \includegraphics{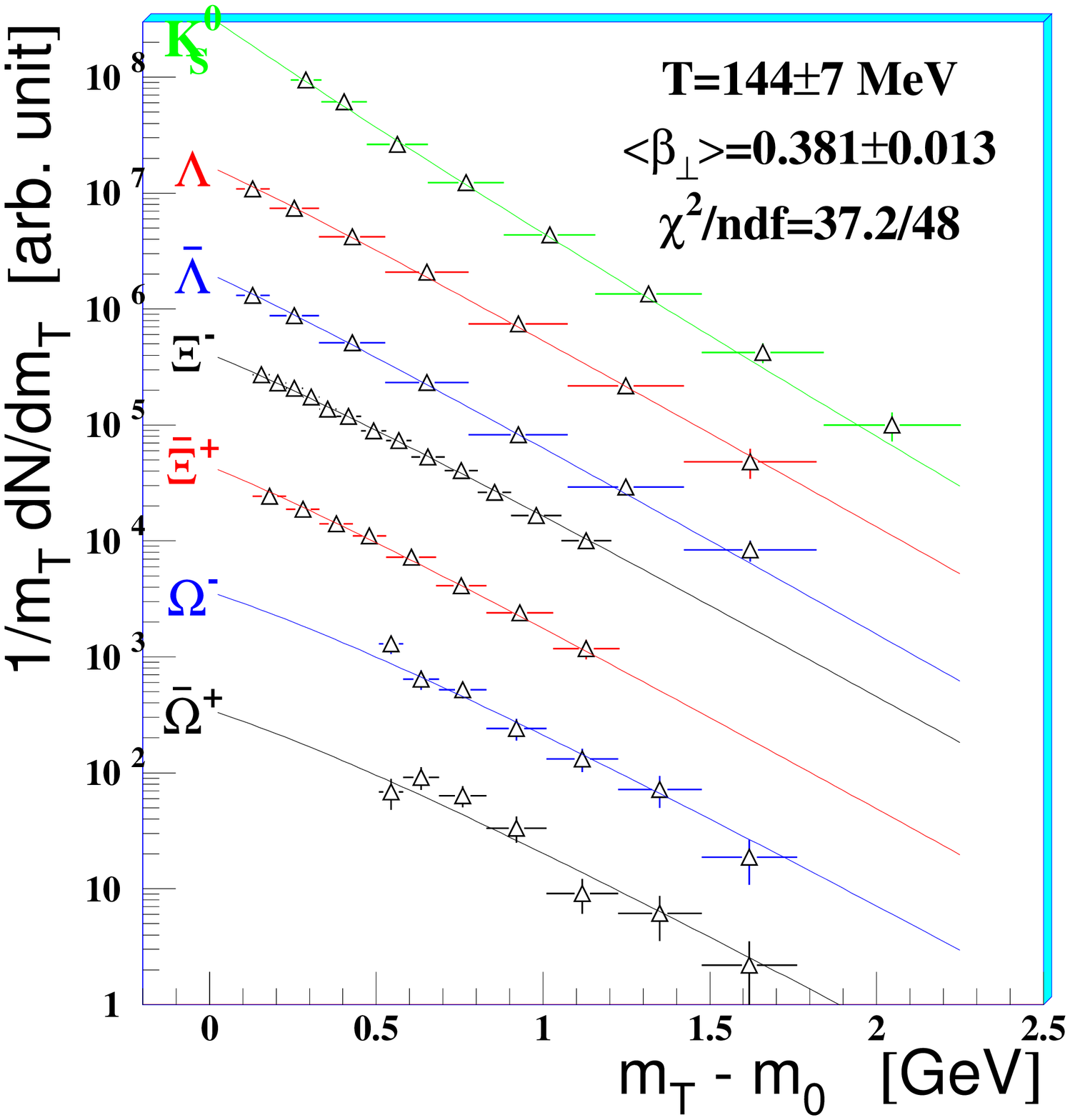}}
  \resizebox{0.45\textwidth}{!}{%
 \includegraphics{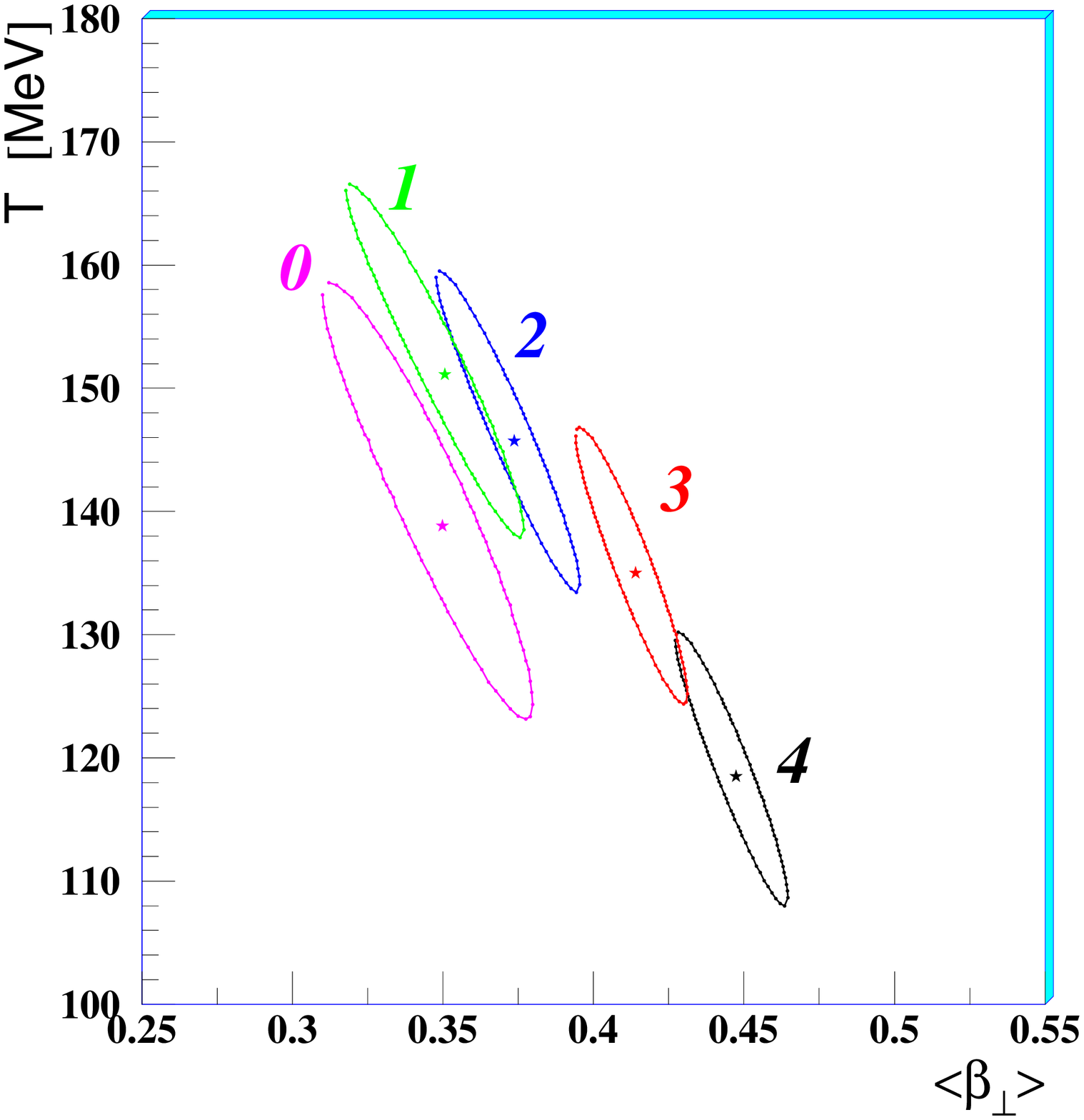}}
\vspace{-0.30cm}
\caption {\label{blast_fit} \small Result of simultaneous fitting of 
\mbox{equation (\ref{blast})} to the hyperon $\mT$ spectra measured in the whole NA57 
centrality range. 
Preliminary $\mathrm{K}^0_{\mathrm{S}}$ data are also included (left). The $1 \sigma$ 
confidence level contours for blast-wave fits to the spectra corresponding to 
individual centrality classes (right).}
\end{figure}

Results of global fitting of the blast-wave model to all strange particle spectra
measured by the NA57 experiment using velocity profiles with $n$ = 0, 1/2 and 1
give similar freeze-out temperatures and average transverse velocities with
adequate statistical significance of fits. In the following, we report the results 
for a linear ($n = 1$) velocity profile. 

As shown in the left portion of figure \ref{blast_fit} it is possible to obtain
a successful description of all particle spectra with common values of the
thermal freeze-out temperature $T$ and average transverse flow velocity 
$\langle \beta_\perp \rangle$. The fitted values of these quantities and their
statistical errors are also shown in figure \ref{blast_fit} \footnote{The systematic 
errors of
quantities $T$ and $\langle \beta_\perp \rangle$ are estimated to be $\pm$14 MeV and
$\pm$0.012 respectively.}.
 
In order to check whether some particle species deviate from the common freeze-out 
description determined according to this model, we have performed separate fitting
of the $\mT$ spectra for singly strange and multi-strange particles (for details see
\cite{qm_04,mt_04}). Results of this analysis show that $\Xi$ hyperons undergo 
a thermal freeze-out which is compatible with that of $\mathrm{K}^0_S$ and $\Lambda$.
The spectrum of $\Omega$ hyperons, although with limited statistics, indicates 
a different behaviour compatible with possible earlier freeze-out.

In order to investigate the centrality dependence of the freeze-out parameters
we have also performed the global fits of $\mT$ spectra for each of the five NA57
centrality classes. The results of these fits are summarized in the right portion of
figure \ref{blast_fit}. Apart from the most peripheral interactions, the centrality
dependence of the $T$ and $\langle \beta_\perp \rangle$ parameters show opposite
trends: the freeze-out temperature is decreasing and the average flow velocity is
increasing with increasing centrality of collision.

\section{Conclusions}
 
The NA57 data on the normalized hyperon yields $E$ at
158 $A$ GeV/$c$ confirm the pattern of strangeness enhancements found by WA97:
the enhancement increases with the strangeness content of the particle reaching 
a factor
$\simeq 20$ for the triply-strange $\Omega$ hyperons. A significant centrality 
dependence of the enhancement for all hyperons and antihyperons (except for the 
$\overline\Lambda$) is observed.

The analysis of the transverse mass spectra in the framework of the blast-wave 
hydrodynamically inspired model shows that $\mT$ spectra 
for all the particles under study can be 
successfully described by common values of the freeze-out temperature and average
transverse velocity. However, the inverse slope of the $\Omega$ hyperon deviates
from the prediction of the blast-wave model tuned to high statistics spectra of
lighter strange particles. A centrality dependence of the 
freeze-out parameters $T$ and $\langle \beta_\perp \rangle$ is observed.

\vfill\eject
\end{document}